\documentclass{article}

\usepackage{arxiv}

\usepackage[utf8]{inputenc} 
\usepackage[T1]{fontenc}    
\usepackage{hyperref}       
\usepackage{url}            
\usepackage{booktabs}       
\usepackage{amsfonts}       
\usepackage{nicefrac}       
\usepackage{microtype}      
\usepackage{lipsum}		
\usepackage{graphicx}
\usepackage{natbib}
\usepackage{doi}
\usepackage{supertabular,lscape,epsfig}
\usepackage{amssymb}
\usepackage{amsmath}
\usepackage{mathtools}
\usepackage{amsthm}

\title{Effect of Radiation Pressure formed at the inner region of accretion disk on the accretion flow in the outer region}

\author{ \href{https://orcid.org/0000-0001-7859-8581}{\includegraphics[scale=0.06]{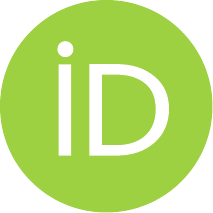}\hspace{1mm}Uicheol Jang}\thanks{Use footnote for providing further
		information about author (webpage, alternative
		address)---\emph{not} for acknowledging funding agencies.} \\
	Department of Astronomy and Space science\\
	Chungnam National university\\
	Daejeon, Korea 34134 \\
	\texttt{temiy@cnu.ac.kr} \\
	\And
    {\includegraphics[scale=0.06]{orcid.pdf}\hspace{1mm}Hongsu Kim} \\
	Center for Theoretical Astronomy\\
	Korea Astronomy and Space science Institute\\
	Daejeon, Korea 34055 \\
	\texttt{chris@kasi.re.kr} \\
	\AND
	Yi Yu \\
	Department of Astronomy and Space science\\
	Chungnam National university\\
	Daejeon, Korea 34134 \\
	\texttt{euyiyu@cnu.ac.kr} \\
}

\renewcommand{\shorttitle}{\textit{arXiv} Template}

\hypersetup{
pdftitle={A template for the arxiv style},
pdfsubject={q-bio.NC, q-bio.QM},
pdfauthor={David S.~Hippocampus, Elias D.~Striatum},
pdfkeywords={First keyword, Second keyword, More},
}

\begin{document}
\maketitle

\begin{abstract}
Studying accretion phenomena is a window into understanding most heavenly bodies, from the birth of stars to Active galactic nuclei(AGN). 
 We would adopt the effect of the radiation pressure which reduces accretion rates$(\dot{M})$ on the accretion phenomena. Shakura-Sunyaev$\alpha$-disk model of disk accretion can be regarded as a good candidate theory of ADAF. Reduction of accreting matter's angular velocity leads to the suppression of the disk luminosity and the surface temperature which essentially implies the transition of the standard accretion disk model from CDAF to ADAF.
\end{abstract}


\section{Introduction}

The accretion phenomena in astrophysics, namely the gravitational capture of ambient matter by compact objects(black hole, neutron star, white dwarf) on stellar scale or the accretion flow onto super massive black holes on galactic scale, certainly, are of major concern for both the theoretical astrophysics community and the observational astronomy community. Most notably, because of the fact that in falling through the steep gravitational potential of a black hole or a neutron star, roughly 10\% of the accreted rest mass energy could be converted into radiation. Namely, the accretion phenomena involve both the gas fluid inflow and the radiation outflow nearly on equal footing. Up until now, however, the standard $\alpha$-disk accretion model of Shakura-Sunyaev(Shakura \& Sunyaev 1973)\cite{1973A&A....24..337S}\cite{1952MNRAS.112..195B} or of Novikov-Thorne(Novikov \& Thorne 1973)\cite{novikov1997black} does not involve in a systematic manner the effect of radiation on the disk accretion phenomena, and the $\alpha$-disk accretion model has been coined, Convection Dominated Accretion Flow(CDAF) as the cooling of accreting gas is efficient. Upon employing the usual hydrodynamics, then, we end up with the familiar physical observable associated with this CDAF model for disk accretion onto various types of systems like:
\begin{itemize}
	\item Young stellar objects(YSO),
	\item Cataclysmic variables(CV),
	\item X-ray binaries(XB), and
	\item Active Galactic Nuclei(AGN)
\end{itemize}

The CDAF model gives the total radiation flux
\begin{equation}
D_c(r)=\frac{3GM\dot{M}}{4\pi r^3}[1-\beta(\frac{r_*}{r})^{1/2}],
\end{equation}
the disk luminosity
\begin{equation}
L_c=\int^\infty_{r_1}4D_c(r)\pi r dr =(\frac{3}{2}-\beta)\frac{GM\dot M}{r_1},
\end{equation}
and the surface temperature
\begin{equation}
T_s(r)=\frac{3GM\dot M}{2\pi b r^3}[1-\beta(\frac{r_1}{r})^{1/2}]^{1/4},
\end{equation}

Where $\dot M$ denotes accretion rate and $\lvert\beta\rvert \le 1 $ and $r_1$ denotes the inner edge of the disk and
 $2Dc(r) = \frac{1}{2} bT^4_s(r)$. Recently, however, observation of some of these systems reveal puzzling anomaly, namely much lower luminosity and the surface temperature than the standard disks associated with the above cited CDAF model. In order to interpret or understand this unexpected anomalous feature of the accretion disks, new underlying models for the accretion flow have been suggested. The most notable, among them, is the Advection Dominated Accretion Flow(ADAF)(Narayan \& Yi 1994\cite{narayan1994advection}, 1995a,b\cite{narayan1995new}; Narayan \& McClintock \& Yi 1996\cite{1996ApJ...471..762A}; Narayan \& Kato \& Honma 1997\cite{narayan1997advection}; Narayan \& Garcia \& McClintock 1997\cite{narayan1997global}; Abramowicz \& Chen \& Kato et al. 1994\cite{abramowicz1994thermal}; Abramowicz \& Chen \& Granath et al. 1996\cite{1996ApJ...471..762A}; Abramowicz \& Lanza \& Percival 1997\cite{abramowicz1997accretion}; Gammie \& Popham 1998\cite{gammie1998advection}; Manmoto \& Mineshige \& Kusunose 1997\cite{manmoto1997spectrum}; Manmoto 2000\cite{manmoto2000advection}) to which we now turn. To briefly summarize the idea or spirit of this new model, it goes as follows. Much lower luminosity and the surface temperature of some disks obviously can be attributed to the inefficient cooling of the disk, which in turn, could be attributed to the situation where the bulk of the liberated thermal energy is carried in by the accreting gas as entropy increase rather than being radiated away. As a consequence, the flow becomes even quasi-spherical. This exciting and quite convincing model, however, lacks its own first principle or even supporting argument thus far despite the active follow-up publications. In the present work, therefore, we would like to put forward the first principle that underlies this ADAF model. Our claim is that, by the systematic inclusion of the radiation pressure on the accretion flow, one can reproduce all the anomalous features of the ADAF.
\section{Modification of "Shakura-Sunyaev alpha-disc model" via the inclusion of the radiation pressure}
In the absence of the radiation force
\begin{align}
	G\frac{m_pM}{r^2}&=m_pa_c=m_pr\Omega^2_c, \\
	\Omega_c&=\left(\frac{GM}{r}\right)^{1/2}.
\end{align}

Where $G$ is the Newton’s gravitational constant; $m_p$ is the mass of the test particle near the central star or black hole(BH); $M$ is the mass of the central star or BH; $r$ is the radial distance of $m_p$ and $M$; $a_c$ is the radial acceleration; $\Omega_c$ is the angular speed of the test particle, in the absence of the radiation pressure. On the other hand, in the presence of the radiation force,

\begin{align}
G\frac{m_pM}{r^2}-\frac{I}{c}\sigma_{T}=m_pr\Omega_A^2, \\
G\frac{m_pM}{r^2}-\frac{L}{c4\pi r^2}\sigma_{T}=m_pr\Omega_A^2, \\
\Omega_A=\left(\frac{1}{r^3}\left\lbrace GM-\frac{\sigma_{T}}{4\pi m_pc}L\right\rbrace\right)^{1/2}.
\end{align}

Where $I$ is the intensity of the central star or BH; $c$ is the speed of light $\sigma_{\textrm{T}}$ is the Thompson scattering cross section; $L$is the luminosity of the central star or BH. $\Omega_A$ is the angular velocity of the test particle, in the presence of the radiation force. Namely, the effect of the radiation pressure on the accretion flow is the reduction of the accreting matter’s angular velocity and eventually the suppression to the total radiation flux, disc luminosity, and surface temperature. To be more concrete, in the absence of the radiation force,

\begin{align}
	D_c(r)&=\frac{3GM\dot{M}}{4\pi r^3}\left(1-\left(\frac{r_*}{r}\right)^{1/2}\right)=2\sigma T^4_s(r), \\
	L_c&=\int^{\infty}_{r_*}2\pi r D_c(r)dr=\frac{GM\dot{M}}{2r_*}.
\end{align}

Total radiation flux, Disc Luminosity, respectively.

\begin{equation}
	T_c(r)=\left(\frac{3GM\dot{M}}{8\pi Gr^3}\left(1-\left(\frac{r_*}{r}\right)^{1/2}\right)\right)^{1/4},
\end{equation}

Where $D_c(r)$ is the total radiation flux; $L_c$ is the disc luminosity; $T_{sc}(r)$ is the surface temperature in the absence of the radiation force; $r_I$ is the inner radius of the disk, and $\omega$ is the surface mass density of the disk, in the absence of the radiation force.
In the presence of the radiation force,

\begin{align}
	D_A(r)&=\frac{3\dot{M}(GM-L\sigma_{T})/4\pi m_pc}{4\pi r^3}\left(1-\left(\frac{r_*}{r}\right)^{1/2}\right)=2\sigma T^4_s(r) \\
	L_A&=\int^{\infty}_{r_*}2\pi r D_A(r)=\dot{M}\left(GM-\frac{\sigma_{T}}{4\pi m_p c}L\right) \\
	T_{sA}(r)&=\left(\frac{3\dot{M}(GM-L\sigma_{T}/4\pi m_p c}{8\pi\sigma r^2}\right)\left(1-\left(\frac{r_*}{r}\right)\right)^{1/4}
\end{align}

Where $L$ denotes the luminosity of the central object and any parameters with subscript A denote corresponding quantities in the presence of the radiation. Namely, they all get suppressed by the reduction of the accreting matter’s angular velocity $\Omega(r)$ via the reduction; renormalization;

\begin{equation}
	GM\rightarrow G\tilde{M}\equiv GM-\frac{\sigma_{T}}{4\pi m_p c}L,
\end{equation}

Obviously, it is interesting to note that as the rotation angular velocity $\Omega_A(r)$, total radiation flux $D_A(r)$, disk luminosity $L_A$ should be positive definite and the surface temperature be real, the luminosity of the central star should be smaller than the Eddington luminosity,

\begin{equation}
	0<L\lesssim \frac{4=pi GMm_p c}{\sigma_{T}}=L_{Edd},
\end{equation}

in the present approach for the modification of the standard accretion disc model by Shakura and Sunyaev(1973) and clearly at the Eddington luminosity $L_{Edd}$, the disc-type accretion process terminates, and the accretion flow makes transition from the disc accretion of Shakura and Sunyaevy(1973) to the spherical, adiabatic, hydrodynamic accretion of Bondi and Hoyle(1952)\cite{1952MNRAS.112..195B}. Particularly, as for the interpretation of the present description as the new candidate theory of ADAF model, it is worthy to note that as the luminosity of the central star approaches the Eddington's critical value $L_{Edd}$, the ADAF nature or characteristic gets more and more ``extremal'' since both the disc luminosity and the surface temperature drop to zero, $L_A$, $T_A(r)\rightarrow 0$. That is, unlike the existing conventional ADAF model, the present approach for the new candidate theory of ADAF model sensitively responds to the luminosity of the central star! Therefore, this last point can be employed as an "observational probe'' to distinguish between our new candidate theory of ADAF and the existing conventional ADAF models such as that of Narayow and YI\cite{narayan1994advection2}. For example, in the existing conventional ADAF models, the inner part of the accretion flow is maintained all the way. provided the central star has low, sub-Eddington luminosity. The essence of the present study and its claim is that the present modification of Shakura-Sunyaev $\alpha$-disk model of disk accretion can be regarded as a new candidate theory of ADAF: reduction of accreting matter’s angular velocity leads to the suppression of the disk luminosity and the surface temperature which essentially implies the transition of the standard accretion disk model from CDAF to ADAF.

\section{Detailed Computations in terms of hydrodynamics}
\subsection{Equations of the radial Structure}
Generally speaking, accretion is the process by which compact objects(BH, NS, WD) or massive stars gravitationally capture or ambient matter. Here we are particularly interested in the disc type accretion of gas onto compact stars of mass $M\sim M_{\odot}$ in the black hole x-ray binaries and onto super massive black holes with $M\simeq 10^6\sim10^9M_{\odot}$ in the active galactic nuclei; (AGN) which are the most likely sources of radiation of energy in the observed rapidly varying emission at high luminosity. Indeed the accretion process onto compact or massive stars is worthy of investigation because in falling through the steep gravitational potential abyss, roughly 10\% of the accreting matter's rest-mass energy may be converted into the radiation. That is, the accretion is a process that can be much more efficient as a cosmic energy source than any other mechanisms in astrophysics such as the nuclear fusion. In practice, we shall consider the ``steady(vanishing partial time derivative)'' accretion of ambient fluid into central stars. We also assume that the fluid would be adiabatic in the first approximation, treating the entropy loss due to radiation being just a small perturbation. Then the fluid will be characterized by an equation of state with the adiabatic index $\Gamma$; $p=\kappa\rho^{\Gamma}$ and thus the speed of sound will be given by
\begin{equation}
	c^2_s=\frac{dp}{d\rho}=\Gamma\frac{p}{\rho}.
\end{equation}
Now, we begin with our findings according to which in the absence of the radiation force, the accretion disc is obviously the Keplerian disc in which the angular velocity(of fluid element) is given by
\begin{equation}
	\Omega_c(r)=\left(\frac{GM}{r^3}\right)^{1/2},
\end{equation}
and its angular momentum is given by
\begin{equation}
	\tilde{L}_e=r^2\Omega_c=(GMr)^{1/2}=\frac{J_c}{M}.
\end{equation}
Meanwhile, in the presence of the radiation force, the accretion disc is still the quasi-Keplerian disc in which the angular velocity is given by
\begin{equation}
	\Omega_A(r)=\left(\frac{1}{3}\left(GM-\frac{\sigma_{T}}{4\pi m_p c}L\right)\right)^{1/2},
\end{equation}
whereas its angular momentum is given by
\begin{equation}
	\tilde{L}_A=r^2\Omega_A=\left(GMr-\frac{\sigma_{T}}{4\pi m_p c}Lr\right)^{1/2}=\frac{J_A}{M},
\end{equation}
namely in both cases, the disc consists of shear flows; having velocity gradients
\begin{align}
	\Omega(r)&>\Omega(r+dr),\\
	\tilde{L}(r)&<\tilde{L}(r+dr).
\end{align}
Therefore, accretion through successive Keplerian orbits forwards the central compact or massive star is possible only if the gas (i.e., the fluid element) constantly lose angular momentum (and transport it out wards) due to some viscous torque.

\subsubsection{Viscous stress and torque}
Therefor, in order eventually to get the expression for this viscous torque, we start with the viscous torque, we start with the viscous force(on viscous stress which essentially would be directly proportional to the shear force or shear stress).
That is,
\begin{equation}
	t_{r\phi}=-2\eta\sigma_{r\phi}
\end{equation}
where $t_{r\phi}$ and $\sigma_{r\phi}$ denote the viscous stress and the shear stress respectively and $\eta$ (g/cm$\cdot$s) denotes some coefficient of dynamic viscosity. Since the shear stress originates from the angular velocity gradient, it would be given by
\begin{equation}
	\sigma_{r\phi}=r\frac{d\Omega}{dr}=-\frac{3}{2}\left(GM-\frac{\sigma_{T}}{4\pi m_pc}L\right)^{1/2}r^{-3/2}=-\frac{3}{2}\Omega_A(r),
\end{equation}
and hence
\begin{equation}
	t_{r\phi}=-\eta\sigma_{r\phi}=\frac{3}{2}\eta\left(\frac{1}{r^3}\left(GM-\frac{\sigma_{T}}{4\pi m_pc}L\right)\right)^{1/2}.
\end{equation}
Namely, this viscous stress $t_{r\phi}$ represents the viscous torque per unit area (around the circumference of the disc layer) exerted in the $\phi$-direction by fluid elements at $r$ on neighboring elements at $r+dr$. Finally, therefore, we are ready to determine the expression for the viscous torque which would be given by
\begin{equation}
	\text{viscous torque}=(\text{viscous force per unit area along $\phi$ direction=viscous stress})\times(\text{area})\times r
\end{equation}
namely,
\begin{align}
	\text{viscous torque}&=\dot{J}^+-\dot{J}^- \\
	&=4\pi r^2ht_{r\phi} \\
	t_{r\phi}4\pi hr^2_I&=\dot{M}\left[\left( GMr-\frac{\sigma_{T}}{4\pi m_pc}Lr\right)^{1/2}-\beta\left( GMr_I-\frac{\sigma_{T}}{4\pi m_pc}r_I\right)^{1/2}\right]
\end{align}
and thus
\begin{equation}
	4\pi r^2_Iht_{r\phi}=(1-\beta)\dot{M}\left(GMr_I-\frac{\sigma_{T}}{4\pi m_pc}Lr_I\right)^{1/2},
\end{equation}
at the inner edge of the disc, $r=r_I$
\paragraph{Note 1}
At this point, it is interesting to note that in the standard Shakura-Sunyaev model, ($\beta=1$) which amounts to the ``zero torque'' inner boundary condition.
\paragraph{Note 2}
Note also the very practical expression for the viscous stress given lastly by
\begin{align}
	t_{r\phi}&=\frac{\dot{M}}{4\pi h}\left(\frac{1}{r^3}\left(GM-\frac{\sigma_{T}}{4\pi m_pc}L\right)\right)^{1/2}\left(1-\beta\left(\frac{r_I}{r}\right)^{1/2}\right)\\
	&=\frac{\dot{M}\Omega_A}{4\pi h}\left(1-\beta\left(\frac{r_I}{r}\right)^{1/2}\right)
\end{align}
which implies that in a steady-state, the viscous stress $t_{r\phi}$ is determined only by the accretion rate $\dot{M}$, the mass $M$, and the luminosity $L$ of the central compact or massive star!
\paragraph{Note 3}
Before we proceed, it might be pedagogically relevant to be more specific on the way we arrive at the expression for the viscous torque given above based essentially on the ``angular momentum conservation.'' As we mentioned earlier for the case at hand where the effect of radiation pressure on the accretion process is included, the inward rate of the angular momentum transport across radius $r$ in the disc due to inflowing fluid is given by
\begin{equation}
	\dot{J}^+=\dot{M}\tilde{L}=\dot{M}\left(GMr-\frac{\sigma_{T}}{4\pi m_pc}Lr\right)^{1/2}.
\end{equation}
Next, if we denote by $\dot{J}^-$, the rate at which the angular momentum is deposited in(last down to) the central compact or massive star through the inner edge of the disc, $r=r_I$ would be given by
\begin{equation}
	\dot{J}^-=\beta\dot{M}\left(GMr_I-\frac{\sigma_{T}}{4\pi m_p c}Lr_I\right)^{1/2}
\end{equation}
for some $\lvert\beta\rvert\leq1$, as the specific angular momentum deposited into the central star cannot exceed the value $\tilde{L}$ at the inner edge of the disc. Lastly, the mass accretion rate $\dot{M}$ would be given as follows essentially from the ``rest mass conservation law''(i.e., the continuity equation):
\begin{equation}
	\frac{\partial\rho_0}{\partial t}+\nabla\cdot\left(\rho_o\vec{u}\right)=0,
\end{equation}
which, upon integration,
\begin{align}
	-\frac{\partial\rho_0}{\partial t}\int d^3x\rho_0&=\int d^3x\nabla\cdot\left(\rho_0\vec{u}\right)\\
	-\frac{dM}{dt}&=\int^{\infty}_{r_I}dr\int^{2\pi}_0rd\phi\int^h_{-h}dz\nabla\cdot\left(\rho_o\vec{u}\right)
\end{align}
yields the equation;
\begin{equation}
	\dot{M}=2\pi rv_r\Sigma,
\end{equation}
where $\Sigma=\int^h_{-h}dz\rho_0$ is the surface mass density.

\subsubsection{Total radiation flux(or surface emissivity) and disc luminosity}
Next, we consider the ``energy conservation'' in the context of standard hydrodynamics. That is, evidently, the heat(or entropy) generated in the accretion disc essentially by the viscosity would be given by, per unit area,
\begin{align}
	\epsilon&=2=eta(\sigma_{ij}\sigma^{ij})\\
	&=2\eta(\sigma_{r\phi}\sigma^{r\phi}+\sigma_{\phi r}\sigma^{\phi r}) \\
	&=4\eta\sigma_{r\phi}\sigma^{r\phi}\\
	&=-4(t_{r\phi}\sigma_{r\phi}).
\end{align}
Therefore, the heat generated by viscosity per unit time, per unit are is given by
\begin{align}
	2h\epsilon&=-8ht_{r\phi}\sigma_{r\phi} \\
	&=\frac{3\dot{M}}{\pi}\Omega^2\left(1-\beta\left(\frac{r_I}{r}\right)^{1/2}\right) \\
	&=\frac{3\dot{M}}{\pi r^3}\left(GM-\frac{\sigma_{T}}{4\pi m_p c}L\right)\left(1-\beta\left(\frac{r_I}{r}\right)^{1/2}\right)
\end{align}
Finally, realizing that the accretion disc possesses two emission surfaces, i.e., the top and bottom surfaces, the total radiation flux(or surface emissivity) can be identified with
\begin{align}
	F_A(r)&=h\epsilon\\
	&=\frac{3\dot{M}}{2\pi r^3}\left(GM-\frac{\sigma_{T}}{4\pi m_p c}L\right)\left(1-\beta\left(\frac{r_I}{r}\right)^{1/2}\right)
	&\equiv Q_e
\end{align}
Note that indeed this is the ``classic'' relationship between the surface emissivity $Q-e$ and the accretion rate $\dot{M}$. Since the disc is assumed to be "thin"(it , the heat(emission) flows out of the disc vertically, rather than radially. Next, the total power radiated, namely the total disc luminosity is given by the integral of the total radiation flux over the entire surface of the disc
\begin{align}
	L_A&=\int^{\infty}_{r_I}2\pi rF_A(r)dr=3\dot{M}\left(GM-\frac{\sigma_{T}}{4\pi m_p c}L\right)\int^{\infty}_{r_I}\left(\frac{1}{r^2}-\beta\frac{r_I^{1/2}}{r^{5/2}}\right)dr \\
	&=\left(\frac{3}{2}-\beta\right)\frac{2\dot{M}}{r_I}\left(GM-\frac{\sigma_{T}}{4\pi m_pc}L\right).
\end{align}
We expect that this total disc luminosity would be the sum of Newtonian gravitational binding energy and rotational kinetic energy extracted from the certain compact or massive object(perhaps via the ``electromagnetic process'', say, of Blanford-Znajek\cite{1977MNRAS.179..433B}), namely,
\begin{equation}
	L=(E_{bind}+E_{rot})\dot{M}
\end{equation}
presumably with
\begin{align}
	E_{bind}&=\frac{1}{r_I}\left(GM-\frac{\sigma_{T}}{4\pi m_p c}L\right), \\
	E_{rot}&=2(1-\beta)\frac{1}{r_I}\left(GM-\frac{\sigma){T}}{4\pi m_p c}L\right).
\end{align}

\subsubsection{surface temperature}
Lastly, assuming that the radiation from the disc is ``black body'' type, the surface temperature of the disc at the radial distance $r$ can be read off as
\begin{align}
	F(r)&=\frac{1}{2}bT_s^4(r) \\
	T_s(r)&=\left(\frac{2F(r)}{b}\right)^{1/4} \\
	&=\left(\frac{3\dot{M}}{\pi b r^3}\left(GM-\frac{\sigma_{T}}{4\pi m_p c}L\right)\left(1-\beta\left(\frac{r_I}{r}\right)^{1/2}\right)\right)^{1/4}
\end{align}
To summarize, therefore, compared with the standard thin disk model of Shakura and Sunyaev or Novikov and Thorne, the important predictions of our new candidate theory of ADAF can be summarized as the suppression of physical characteristics of the accretion disc;
\begin{align}
	\frac{L_A}{L_c}&=1-\frac{\sigma_{T}}{4\pi m_pc}\frac{L}{GM}\equiv1-f, \\
	\frac{T_s^A(r)}{T^c_s(r)}&=\left(1-\frac{\sigma_{T}}{4\pi m_p c}\frac{L}{GM}\right)^{1/4}\equiv(1-f)^{1/4},
\end{align}
and hence more concretely,
\begin{align}
	T^A_s(r)&\simeq(3\times 10^7\text{K})\left(\frac{\dot{M}}{10^{-9}M_{\odot}\text{/year}}\right)^{1/4}\left(\frac{M}{M_{\odot}}\right)^{-1/2}\left(\frac{GM}{r}\right)^{3/4} \\
	&\times\left(1-\beta\left(\frac{r_I}{r}\right)^{1/2}\right)^{1/4}\left(1-\frac{\sigma_{T}}{4\pi m_pc}\frac{L}{GM}\right)^{1/4}.
\end{align}
\paragraph{Note 1}
That is, the total disc luminosity responds very sensitively to the underlying disc model and thus gets suppressed noticeably whereas the surface temperature responds insensitively and remains nearly the same! Apparently this feature of our new candidate theory of ADAF is consistent with the typical characteristic of the standard, existing, conventional ADAF model, say, of Narayan and Yi according to which, the temperature of the accreting gas is nearly Virial where as most of the viscously dissipated energy(i.e., the radiation) is trapped or stored as entropy in the accreting gas and is eventually advected into the central star or black hole leading to the low luminosity, abnormal accretion flow.
\paragraph{Note 2}
Our new candidate theory of ADAF also appears to allow us to uncover or unveil the (possibly) true nature of the advection energy as, following the terminology of the existing conventional ADAF model, the fraction of the viscously dissipated energy which is being radiated is given by $(1-f)L_c$ whereas the fraction $fL_c$ which is being stored in the accreting gas and is eventually being advected into the central star or black hole originates from the recoil of the accreting gas by the scattering of the radiation from the central star or black hole. That is, according to our new candidate theory of ADAF, it is the radiation pressure of the central star or black hole (and the recoil of accreting gas due to it) which renders part of the viscously dissipated energy to be advected into the central star/black hole!
Therefore if we restrict our interest to the  context of our new candidate theory of ADAF, the ``advection dominated accretion flow'' limit, $f\rightarrow 1$ clearly corresponds to the case when the central object is maximally radiating at the critical Eddington luminosity (in which case the accretion flow becomes essentially quasi-spherical as we shall see in a moment)whereas just the opposite, ``little advection'' limit $f\rightarrow 0$ amounts to the case when the central object possesses very low luminosity.
\paragraph{Note 3}
Lastly, we provide the emission spectrum of the accretion disc in some systems of our interset predicted by our new candidate theory of ADAF.
\begin{itemize}
	\item for  Black hole x-ray binaries
	\begin{align}
		T_s(r)&\simeq (10^7 \text{K})\left(\frac{\dot{M}}{10^{-8}M_{\odot}\text{/year}}\right)\left(\frac{M}{3M_{\odot}}\right)^{-1/2}\left(\frac{10GM}{r}\right)^{3/4}\\
		&\times\left(1-\frac{\sigma_{T}}{4\pi m_p c}\frac{L}{GM}\right)^{1/4}
	\end{align}
	\item for Active Galactic Nuclei(AGN)
	\begin{align}
		T_s(r)&\simeq(10^5\text{K})\left(\frac{\dot{M}}{M_{\odot}\text{/years}}\right)\left(\frac{M}{10^8M_{\odot}}\right)^{-1/2}\left(\frac{10GM}{r}\right)^{3/4}\\
		&\times\left(1-\frac{\sigma_{T}}{4\pi m_p c}\frac{L}{GM}\right)^{1/4}
	\end{align}
\end{itemize}

\subsection{Equation for the description of bipolar outflows/jets: The radial component of the momentum equation(Euler equation)}
Net, in the context of our new candidate theory of ADAF in which the effect of radiation pressure on the accretion process is included, we would like to look for a natural description/explanation for the widespread occurrence of bipolar outflows/jets that are so ubiquitous in accreting systems such as, say, the black hole x-ray binaries, micro-quasars, and the active galactic nuclei(AGNs).
To this end, we start with the general form of the momentum conservation equation or Euler equation
\begin{equation}
	\rho\left(\frac{\partial}{\partial t}+\vec{v}\cdot\nabla\right)\vec{v}=-\nabla p-\rho\nabla\Phi_N+\vec{F}_{rad}+\eta_v\left(\nabla^2\vec{v}+\frac{1}{3}\nabla(\nabla\cdot\vec{v})\right)
\end{equation}
where the alst term on the right hand side is the usual viscous force with $\eta_v$ denoting the (microscopic) kinematic viscosity and
\begin{equation}
	F_{rad}=\frac{I}{c}\sigma_{T}
\end{equation}
is the radiation force that we systematically include in our present new theory of ADAF. In the present treatment below, we shall neglect the viscous force term and attempt to solve the radial and (possibly) vertical components of this momentum conservation equation for a solution representing the bipolar outflow/jet. To be more concrete, our strategy to achieve this goal is as follows.
In order to demonstrate actual occurrence of the bipolar outflow/jet in our new candidate theory of ADAF, we should find a solution to the radial and (possibly) vertical components of this momentum conservation equation that turns the inflowing gas around and renders it flow outward on a radial trajectory to reach infinity with a net positive energy. In the standard thin disk model of Shakura and Sunyaev or Novikov and Thorne and even in the existing conventional ADAF model, the only force apart form the viscous force involved in the hydrodynamics equations (including this momentum conservation equation) is the attractive Newtonian gravity exerted by the central subject and hence the generic motion of the accreting gas is apparently radial inflow which essentially makes the construction of the outflow mechanism very difficult. In our new candidate theory of ADAF, on the other hand, we have, in this Euler equation, the repulsive radiation force in addition to the attractive Newtonian gravitational force. Interestingly, therefore, if the repulsive radiation force overwhelms the attractive Newtonian gravity in their competition, one would be able to find a solution to the hydrodynamics equations (including this momentum conservation equation) which demonstrate the occurrence of outflows. Namely, the repulsive radiation force clearly plays the key role of scattering/reflecting off the incident accreting gas-plasma!
\subsection{Equations of the vertical structure}
Since normally there is no net motion of the gas/fluid elements in the vertical direction of the accretion disc, momentum conservation along z-direction reduces to a ``hydrostatic equilibrium'' condition.
Consider now the vertical component of the momentum conservation equation(Euler equation) given earlier. The physical meaning and the role of this equation is as follows. By equating the component of the (Newtonian) gravitational force plus that of the radiation force of the central compact or massive star along $\hat{e}_z$ to the vertical pressure gradient in the disc one can determine the thickness of the disc and the vertical mass density profile of the accretion disc.

viz.
\begin{equation}
	\rho\left(\frac{\partial}{\partial t}+\vec{v}\cdot\nabla\right)\vec{v}\cdot\hat{e}_z=\left(-\nabla p-\rho\nabla\tilde{\Phi}\right)\cdot\hat{e}_z,
\end{equation}
where $-\nabla\tilde{\Phi}=-\nabla\Phi_N+\vec{F}_{rad}$ is the net, effective gravitational force including the radiation force, namely,
\begin{equation}
	\tilde{\Phi}=\frac{1}{r}\left(-GM+\frac{\sigma_{T}}{4\pi m_pc}L\right).\\
\end{equation}
Then
\begin{align}
	\frac{\partial v_z}{\partial t}+\vec{v}\cdot\nabla v_z&=-\frac{1}{\rho}\frac{dp}{dz}-\frac{d}{dz}\left(-\frac{GM}{r}+\frac{\sigma_{T}}{4\pi m_pcr}L\right),\\
	\frac{1}{\rho}\frac{dp}{dz}&=-\left(\frac{z}{r^3}\right)\left(GM-\frac{\sigma_{T}}{4\pi m_pc}L\right),
\end{align}
and using $c_s^2=\frac{dp}{d\rho}$,
\begin{align}
	\frac{c^2_s}{\rho}\frac{d\rho}{dz}&=-z\Omega^2_z, \\
	\int\frac{d\rho}{\rho}&=-\frac{\Omega_A^2}{c^2_s}\int zdz,\\
	\ln\rho&=-\frac{\Omega^2_A}{2c^2_s}z^2+\text{const},\\
	\frac{\rho(z)}{\rho_0}&=e^{-\frac{\Omega_a^2}{2c_s^2}z^2}.
\end{align}
Thus the vertical mass density profile of the accretion disc is given by the Gaussian function,
\begin{equation}
	\rho(z)=\rho(z=0)\exp\left[-\frac{z^2}{2h^2}\right],
\end{equation}
where so called ``scale heigh''
\begin{align}
	h&\equiv H/R\\
	&=c_s/\Omega_A\\
	&=\left(\frac{p}{\rho}\right)^{1/2}\left(\frac{GM}{r^3}-\frac{\sigma_{T}}{4\pi m_pc}L\right)^{-1/2}
\end{align}
estimates the ``half thickness'' of the disc. Here the ``sound speed' is normally assumed to be
\[
c_s=\left(\frac{dp}{d\rho}\right)^{1/2}=
\begin{dcases}
\frac{p^{gas}}{\rho}=\frac{T}{m_p}& \text{\quad(whole disc)},\\
\frac{p^{rad}}{\rho}=\frac{1}{3}b\frac{T^4}{\rho}              & \text{\quad (inner disc)}.
\end{dcases}
\]
Finally, it is interesting to note that as the luminosity $L$ of the central object gets higher, the Keplerian angular velocity of the fluid element gets smaller and as a result, the scale height gets larger; i.e., the disc becomes thicker. Particularly, when the luminosity of the central compact or massive star approaches the critical Eddington value,
\begin{equation}
	L\rightarrow L_{Edd}=\frac{4\pi Gm_pc_M}{\sigma_{T}},
\end{equation}
the Keplerian angular velocity of the accreting gas drops to zero and the thickness of the accretion disc blows up!
Clearly this point implies that our new candidate model theory of ADAF predicts that if the central object is maximally-radiating at the Eddington luminosity, the accretion flow is essentially quasi-spherical and generically, as the central object gets more and more luminous, the accretion flow makes transition from the disc-type toward the spherical Bondi-type\cite{1952MNRAS.112..195B}.

\section{Class of ADAF models to account for low-luminosity abnormal accretion flow observed in some black hole X-ray binaries and AGN}
\begin{itemize}
	\item[(1)] Optically thick ADAF-model(Katz, 1977\cite{katz1977x}; Begelman, 1978\cite{1978A&A....70..583B}; Begelman and Meier, 1982\cite{begelman1982thick}; Abramowicz, et al., 1988\cite{abramowicz1988slim}) At super Eddington accretion rates/luminosity, the inflowing accreting gas is optically thick and hence traps(captures) most of the radiation(i.e., the viscous energy) instead of being radiated and carries it inward or ``advects'' it onto the central star/black hole resulting in the low luminosity abnormal accretion flow inconsistently with the standard thin disc model first developed by Shakura and Sunyaev, 1973\cite{1973A&A....24..337S}; Novikov and Thorne, 1973\cite{novikov1997black}; Lynden-Bell and pringle, 1974\cite{lynden1974evolution}.
	\item[(2)]Obtically thin ADAF-model(Ichimaru, 1973\cite{1973A&A....24..337S}; Rees et al., 1982\cite{rees1982ion}; Narayan and Yi; 1994\cite{narayan1994advection3}, 1995\cite{narayan1995new}; Abramowicz, et al., 1994\cite{abramowicz1994thermal}) In the opposite limitof low, sub Eddington accretion rates/luminosity, on the other hand, the accreting gas has a very low density and thus is optically thin and is unable to cool efficientyl withing and accretion time. The radiation (i.e., the viscous energy) is staored in the inflowing gas as thermal energy and is ``advected'' into the central star/black hole leading to the low luminosity abnormal accretion flow in contradiction to the standard thin disc model of Shakura and Sunyaev.
	\item[(3)] Our new candidate theory of ADAF
	At a generic accretion rate/luminosity, anywhere between the low sub-Eddington accretion rates/luminosity and that of super Eddington, it is demonstrated that one can account for the low luminosity abnormal accretion flow even in the context of the standard thin disc model of Shakura and Sunyaev or of Novikov and Thorne\cite{novikov1997black}, provided one include the effect of radiation pressure on the accretion process in a simple but straight forward manners.
	\end{itemize}
	
\section{Concluding Remarks}
To conclude, the lessons we learned from the study of our new candidate theory of ADAF which, among others, includes the effect of radiation pressure on the accretion process can be summarized as follows:
\begin{itemize}
	\item[(1)] For the case when the central star is a non luminous black hole, regardless of whether it is non-rotating or rotating(posseising the well known frame dragging effect) in practice, the stationary adiabatic accretion flow is most likely of disc type exhibiting features which would faithfully reflect the predictions of the standard thin disc model of Shakura and Sunyaev or Novikov and Thorne(although the spherical, adiabatic and hydrodynamical accretion of Bondi-Hoyle type might be the case, in principle).
	\item[(2)] For the other case when the central star is highly luminous exhibiting nearly critical or even super Eddington luminosity. According to the results of the study of our new candidate theory of ADAF, the accretion flow would exhibit Features that deviates (or departs) significantly from the predictions of the standard thin disc model of Shakura and Sunyaev. That is, basically due to the effect of the radiation pressure on both the radial and vertical structures of the disc, the low luminosity abnormal accretion flow of nearly spherical type is very likely to take place and even for the case when the central star possesses the generic accretion rate or luminosity(well below Eddington vertical value) the low low luminosity abnormal accretion flow of nearly spherical type is has a good chance to happen explaining the recent observations in some black hole x-ray binaries and galactic nuclei.
\end{itemize}

\appendix
\section{An analytical solution based on the self similar solution of Spruite et. al. (1987)}
Fortunately, an exact self-consistent solution to the hydrodynamics equations (including, of course the Euler equation) in the context of our new candidate theory of ADAF representing such bipolar outflows/jets is indeed available. it can be derived from the by-now well known self-similar solutions discovered first by Spruite et. al. (1987)\cite{spruit1987spiral} and then elaborated later on by Narayan. R., and Yi. I.,(1994, 1995)\cite{narayan1994advection},\cite{narayan1994advection2} in their construction of the optically thin ADAF model.
Therefore in the following we shall briefly describe its construction. As a typical approach to the study of a steady axisymmetric accretion flow in the standard theory of thin accretion disks, we start with the vertical average of the flow to consider a two-dimensional accretion flow in the equatorial $r\phi$ plane described by the following height-integrated hydrodynamics equations consisting of the continuity equation, the radial and azimuthal components of the momentum conservation equation(i.e, the Euler equation) and the energy conservation equation;
\begin{align}
	\frac{d}{dr}(\rho r Hv)&=0,\\
	v\frac{dv}{dr}-r\Omega^2&=-r\Omega^2_K-\frac{1}{\rho}\frac{d}{dr}(\rho c^2_s),\\
	v\frac{d(r^2\Omega)}{dr}&=\frac{1}{\rho rH}\frac{d}{dr}\left(\frac{\alpha\rho c^2_sr^3H}{\Omega_K}\frac{d\Omega}{dr}\right)\\
	\rho vHT\frac{ds}{dr}&=q^+-Q^-\\
	&=\frac{2\alpha\rho c^2_sr^2 H}{\Omega_K}\left(\frac{d\Omega}{dr}\right)^2-Q^- \\
	&=f\frac{2\alpha\rho c^2_sr^2H}{\Omega_K}\left(\frac{d\Omega}{dr}\right)^2,
\end{align}
where $\Omega_K=\left(\frac{G\tilde{M}}{r^3}\right)^{1/2}$, $v_k=r\Omega_K=\left(\frac{G\tilde{M}}{r}\right)^2$, $c_s=P/\rho$ are the Keplerian angular velocity, and the isothermal sound speed respectively and the surface (mass) density of the gas $\Sigma$ is given by $\Sigma=2\rho H$ with $H$ being the so called scale height, $H=c_s/\Omega_K$. We assumed the so called $\alpha$-disc prescription by Shakura and Sunyaev(1973)\cite{1973A&A....24..337S} in which the kinematic coefficient of viscosity is given by
\begin{equation}
	\nu=\alpha c_{stt}=\frac{\alpha c^2_s}{\Omega_K}.
\end{equation}
Next, the parameter $f$ measures the degree to which the flow is advection dominated so that the two extreme limits $f=1$ and $f=0$ represent respectively, the pure advection(i.e., no radiative colling) and the very efficient colling. For convenience, we defined a parameter $\epsilon\equiv(5/3-\gamma)/(\gamma-1)$ and $\epsilon'\equiv\epsilon/f$ with $\gamma$ denoting the ration between specific heats and obviously $\epsilon=0$ in the limit $\gamma=5/3$ and $\epsilon=1$ when $\gamma=4/3$. For simplicity, we assumed that $\epsilon'$
 is independent of $r$. In the energy conservation equation, $Q^+$ denotes the energy input per unit area due to the viscous dissipation while $Q^-$ denotes the energy loss through radiative colling. As is well known by now, it has been known for some time that these four height-averaged differential equations admit a self-similar solution(references) of the form
 \begin{equation}
 	\rho\propto r^{-3/2},v\propto r^{-1/2}, \Omega\propto r^{-3/2}, c^2_s\propto r^{-1},
 \end{equation}
 or more precisely,
\begin{align}
	v&=-\frac{5+2\epsilon'}{3\alpha}g(\alpha,\epsilon')v_k\approx-\frac{3\alpha}{5+2\epsilon'}v_K, \\
	\Omega&=\left(\frac{2\epsilon'(5+2\epsilon')}{9\alpha^2}g(\alpha,\epsilon'\right)^{1/2}\Omega_K\approx\left(\frac{2\epsilon'}{5+2\epsilon'}\right)^{1/2}\Omega_K,\\
	c^2_s&=\frac{2(5+2\epsilon')}{9\alpha^2}g(\alpha,\epsilon')v^2_K\approx\frac{2}{5+2\epsilon'}v^2_K,\\
\end{align}
where
\begin{align}
	g(\alpha,\epsilon')&\equiv\left(1+\frac{18\alpha^2}{(5+2\epsilon)^2}\right)^{1/2}-1, \\
	\epsilon&\equiv\frac{5/3-\gamma}{\gamma-1},\\
	\epsilon'&=\frac{\epsilon}{f},\\
	\frac{4}{3}&\lesssim\gamma\leq\frac{5}{3}.
\end{align}
Note that the positivity of the Bernoulli constant $B_e$ and hence $b$ implies that the viscous stress transfers energy from small to large radii. Thus whenever $b$ is positive, it means that, if one could somehow turn the inflowing gas around to let it flow outward on a radial trajectory, the gas would reach infinity with a net positive energy.
It is noteworthy that the volume mass density $\rho$ maybe obtained in practice from the mass accretion rate via $\dot{M}=2\pi rv\Sigma=4\pi rvH\rho$.
Obviously in the limit of very efficient colling $f\rightarrow 0$, $\epsilon'\rightarrow\infty$, the solutions given in equations ... reduce to those of the standard thin disc model(In this paper, we don't consider Thick disk case\cite{jang2020thick}) with $v,c_s\ll v_K$ and $\Omega\rightarrow\Omega_K$ as they should. Thus in the present work we are mainly interested in the opposite limit of advection dominated accretion flow where $f$ is a reasonable fraction of unity; namely, where it has a generic value and thus $\epsilon'\sim\epsilon<1$.
Note now that the radial component of the momentum conservation equation indicates that the self-similar solution turns out to satisfy the interesting relation;
\begin{equation}
	\frac{1}{2}v^2+r^2\Omega^2-\Omega^2_K+\frac{5}{2}c^2_s=0,
\end{equation}
which in turn allows us to compute the normalized parameter $b=B_e/v_k^2$ where $B_e$ is the well known Bernoulli constant, in adiabatic flows in the absence of viscosity;
\begin{align}
	b&=\frac{1}{v_K^2}\left(\frac{1}{2}v^2+\frac{1}{2}r^2\Omega^2-r^2\Omega^2_K+\frac{\gamma}{\gamma-1}c^2_s\right)\\
	&=\frac{r^2\Omega^2}{2v^2_K}+\left(\frac{\gamma}{\gamma-1}-\frac{5}{2}\right)\frac{c_s^2}{v^2_K}\\
	&=\frac{(3f-1)\epsilon'}{5+2\epsilon'}.
\end{align}

 \section{Some basics in classical theory of radiation}
central star$=$ a ``black'' body

accreting object $=$ an atom (with cross-ectional area $\sigma_{T}$)

luminasity=power
is the radiated energy per unit time mostly via an emission of (massless) photons moving at the speed of light $c$.
\begin{equation}
	E=pc;
\end{equation}
$\text{power}=L=\frac{\Delta E}{\Delta t}=c\left(\frac{\Delta P}{\Delta t}\right)=c F$.
Intensity is the radiated (transferred) energy per unit time per unit area;
\begin{equation}
	I=\frac{\Delta E}{A\Delta t}=\frac{1}{A}\left(\frac{\Delta E}{\Delta t}\right)=\frac{Power}{A}=\frac{Power}{4\pi r^2}
\end{equation}
Radiation pressure : Radiation stress tensor
\begin{equation}
	P_{rad}=\frac{F}{A}=\frac{1}{A}\left(\frac{\Delta P}{\Delta t}\right)=\frac{1}{A\Delta t}\left(\frac{\Delta E}{c}\right)=\frac{I}{c}
\end{equation}
For perfectly absorbing object $\left(F=\Delta p/\Delta t\right)$ $P_{rad}=I/c$

For perfectly reflecting object $\left(F=2\Delta p/\Delta t\right)$ $P_{rad}=2I/c$

radiation force. Lastly the ``radiation force'' on an object with the effective cross-sectional area $\sigma$ is
\begin{equation}
	F_{rad}=\sigma P_{rad}.
\end{equation}
Thus if the accreting object is the ``free-electron'', its (classical) cross section is $\sigma_{T}$ and hence
\begin{equation}
	F_{rad}=P_{rad}\sigma_{T}=\frac{I}{c}\sigma_{T}.
\end{equation}

\bibliographystyle{unsrtnat}
\bibliography{mybibfile}  






\end{document}